\documentclass[twocolumn,prl,showpacs,floatfix]{revtex4}
\usepackage{graphicx}
\usepackage{bm}

\newcommand{\toCA}[1]{}

\newcommand{\be}{\begin{equation}}
\newcommand{\ee}{\end{equation}}
\newcommand{\ba}{\begin{eqnarray}}
\newcommand{\ea}{\end{eqnarray}}

\begin{document}

\title{Nonmonotonic behavior of resistance in 
a superconductor-Luttinger liquid junction}
\author{H.-W. Lee,$^{1,2}$ Hyun C. Lee,$^3$, Hangmo Yi,$^1$ 
and Han-Yong Choi$^3$}
\affiliation{$^1$School of Physics, Korea Institute for Advanced Study, 
207-43 Cheongryangri-dong, Dongdaemun-gu, Seoul 130-012, Korea}
\affiliation{$^2$School of Physics and Physics Research Division,
Seoul National University, Seoul 151-747, Korea}
\affiliation{$^3$BK21 Physics Research Division 
and Institute for Basic Science,
School of Physics, Sung Kyun Kwan University,
Suwon 440-746, Korea}
\date{\today}

\begin{abstract}
Transport through a superconductor-Luttinger liquid junction is considered. 
When the interaction in the Luttinger liquid is repulsive,
the resistance of the junction with a sufficiently clean interface 
shows nonmonotonic temperature- or voltage-dependence 
due to the competition between
the superconductivity and the repulsive interaction.
The result is discussed in connection with
recent experiments on single-wall carbon nanotubes
in contact with superconducting leads.
\end{abstract}

\maketitle


In one dimensional systems, it is believed~\cite{Voit94RPP} that
the usual Fermi liquid description breaks down
and systems become a Luttinger liquid (LL).
LL behavior has been predicted~\cite{Egger97PRL} 
for single-wall carbon nanotubes (CNT),
and demonstrated in experiments~\cite{Bockrath97Science}.
Recently hybrid systems, 
which consist of LLs and superconductors (S),
have also received theoretical attention~\cite{Fazio95PRL,%
Maslov96PRB,Takane96JPSJ,Affleck00PRB,Caux01cond-mat}.
A couple of recent experiments~\cite{Kasumov99Science,Morpurgo99Science}
reported transport measurements on CNTs 
in contact with superconductors.

The contact with superconductors may generate interesting phenomena.
In a S--normal metal (N) junction,
the junction resistance may be reduced due to 
the proximity effect mediated by the Andreev process~\cite{Blonder82PRB}.
This effect is more evident in a junction with a clean interface.
In a S-LL junction, an interesting competition may arise
between the superconductivity and the electron-electron interaction
since the repulsive interaction in a LL tends to
enhance the resistance~\cite{Kane92PRB}.
The competition was investigated in various theoretical works.
However previous studies were mostly focused either  
on the tunneling regime~\cite{Fazio95PRL},
where the Andreev process is strongly suppressed,
or
on the low energy regime~\cite{Maslov96PRB,Takane96JPSJ,Affleck00PRB} 
far below the superconducting gap $\Delta$,
where the repulsive interaction always wins over the superconductivity.

\begin{figure}
\includegraphics[width=0.80\columnwidth]{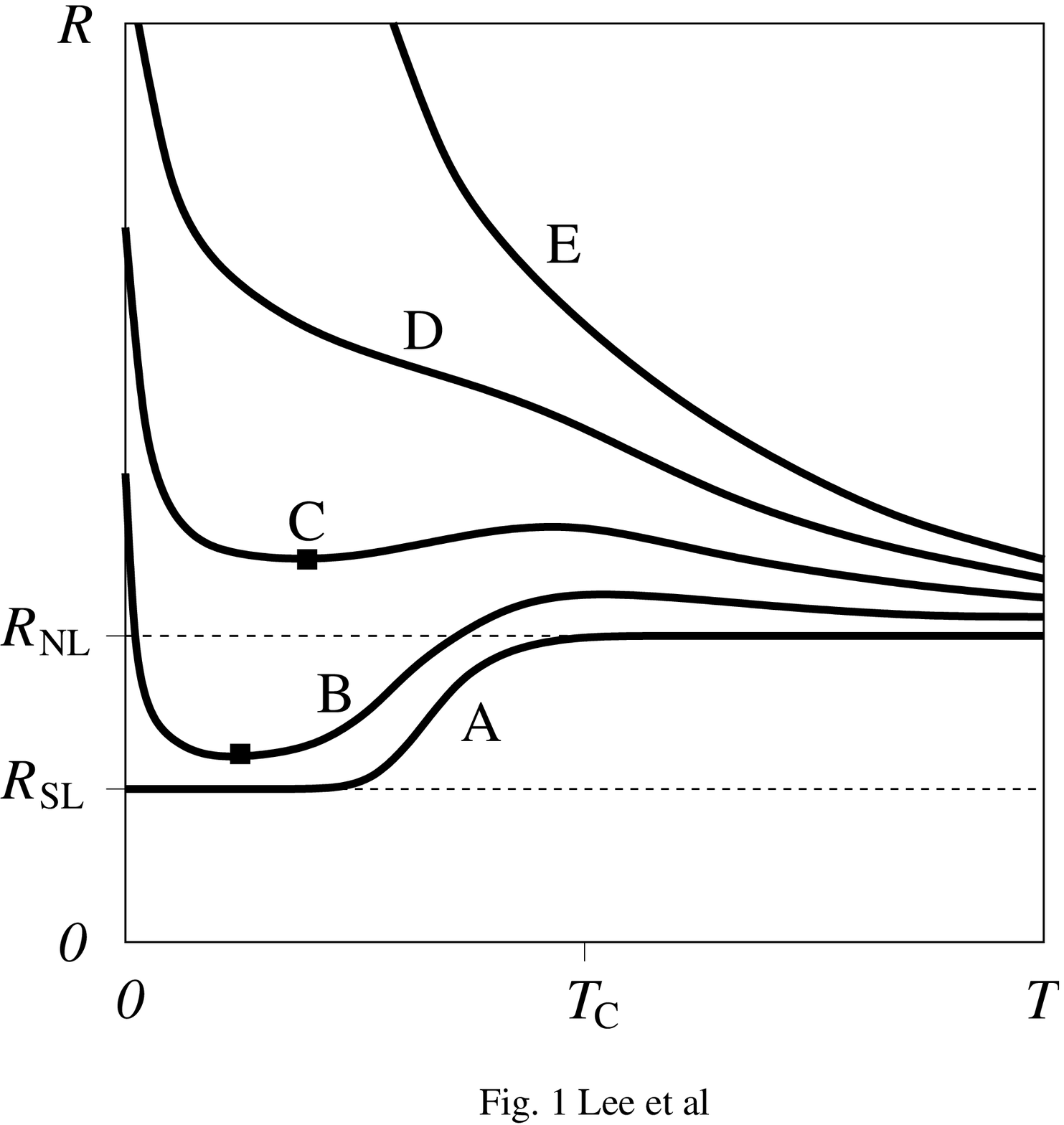}
\caption{
A schematic diagram of the zero bias resistance $R$
as a function of the temperature $T$.
The curve A shows $R(T)$ for an ideal interface ($U_0=0$).
The curves B, C, and D describe 
progressively less ideal interfaces,
and the curve E shows $R(T)$ for a poorly transmitting interface (large $U_0$).
In the curves B and C, $R(T)$ below $T_c$ is nonmonotonic
due to the competition between the superconductivity and 
the repulsive interaction.
The black squares in the curves B and C denote
the crossover temperature $T_{\rm cr}$.
$R_{\rm NL}=G_{\rm NL}^{-1}$ and $R_{\rm SL}=G_{\rm SL}^{-1}$ (see text).
}
\label{RG_flow}
\end{figure}

\begin{figure}
\includegraphics[width=0.85\columnwidth]{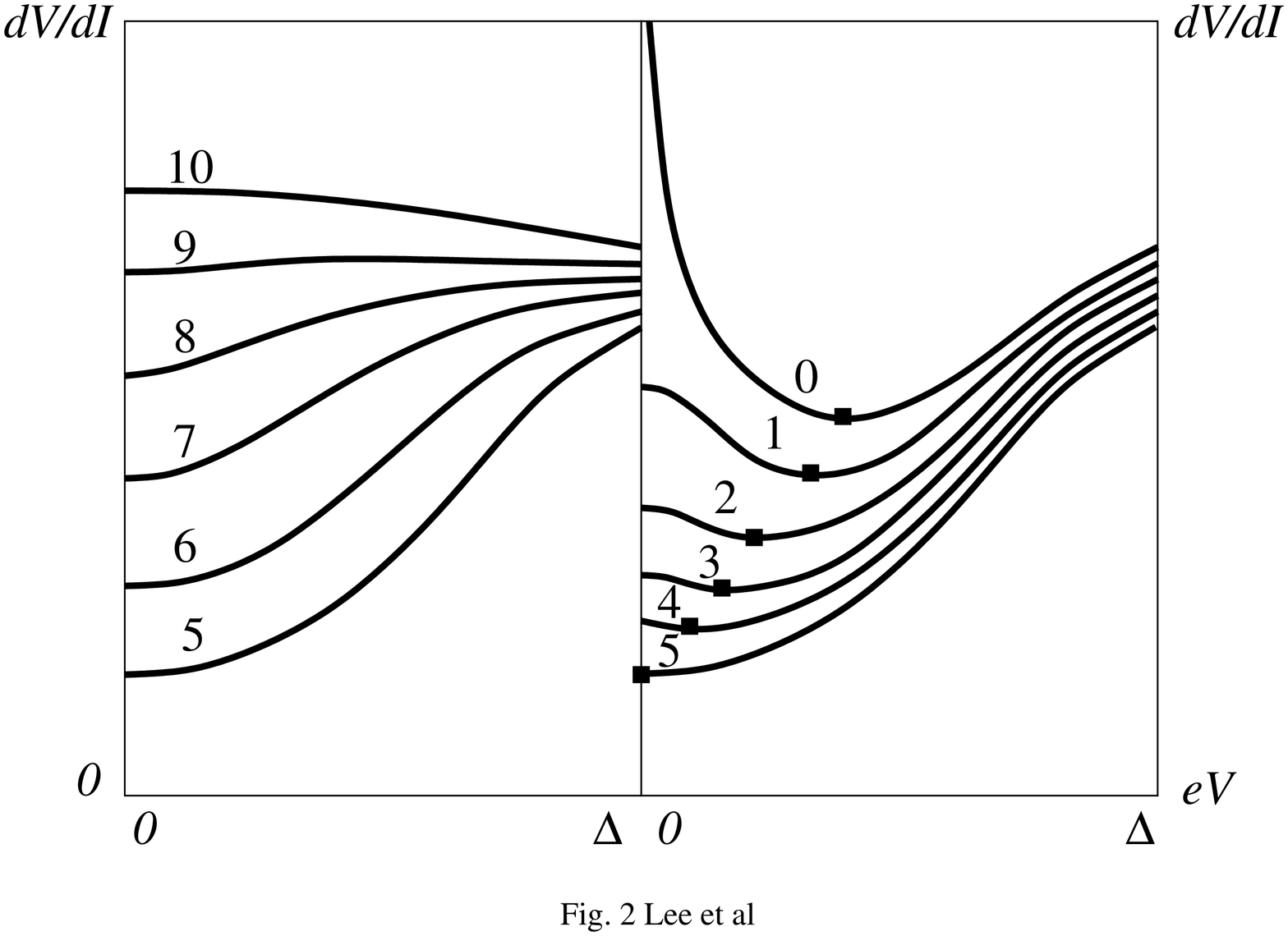}
\caption{
A schematic drawing of the $dV/dI$--$V$ characteristic
for an almost ideal interface.
$T\approx T_c$ for the curve 10 and
the temperatures for the curves 9, 8, \ldots, 1
are progressively lower.
The curve 0 denotes the characteristic
at $T=0$.
Curves are vertically shifted for the clarity.
Without the shift, all curves should almost collapse over each other 
near $eV= \Delta$.
When $T$ is below the crossover temperature $T_{\rm cr}$
(temperature for the curve 5),
the characteristic becomes nonmonotonic in $V$
due to the competition between the superconductivity and
the repulsive interaction.
The black squares in the curves 0-5 denote
the crossover voltage $V_{\rm cr}(T)$. 
}
\label{R-V}
\end{figure}

In this paper, we investigate the competition
in the energy range $\sim \Delta$
for a S-LL junction with a clean interface.
We first examine two extreme energy regimes,
far below $\Delta$ and far above $\Delta$,
and interpolate these results to the intermediate energy regime.
Although quantitative predictions cannot be obtained with this approach, 
we still find that qualitative behaviors at the energy scales $\sim \Delta$
can be determined without ambiguity.
In particular, it is found that
the competition between the superconductivity and
the interaction results in 
a nonmonotonic behavior of resistance~(Figs.~\ref{RG_flow} and \ref{R-V})
when the junction interface is sufficiently clean.
The result is discussed in connection with
a recent experiment~\cite{Morpurgo99Science}.

{\it Model}:  A purely one-dimensional 
model of a S-LL junction is used,
where the superconductor is also modeled as a one-dimensional conductor 
with the superconducting potential $\Delta$~\cite{Affleck00PRB}.
We choose the S to reside in the negative
semi-infinite space $x<0$ and the LL in
the positive semi-infinite space $x>0$.
The bosonization formula~\cite{Voit94RPP},
\be
\psi_{r,\sigma} \! \propto
\exp\left[ irk_Fx \! +i(r\theta_c \! + \! r\sigma \theta_s \! + \!
\phi_c \! + \! \sigma \phi_s)
/\sqrt{2}\right],
\label{eq:bosonization}
\ee
is used,
where $r=\pm 1$ for R (right)/L (left) and $\sigma=\pm 1$ for spin up/down.
The bosonic fields satisfy the commutation relation
$[\theta_\mu(x),\partial_y \phi_{\mu'}(y)]
=- i\pi \delta(x-y)\delta_{\mu\mu'}$, where $\mu,\;\mu'=$ c (charge) 
or s (spin).
We study the bosonized Hamiltonian $H=H_{LL}+H_{S}+H_{\rm imp}$.
$H_{LL}$ is the usual LL Hamiltonian,
\begin{equation}
H_{LL}\! =\! \! \!   \sum_{\mu=c,s}\int_0^\infty \! \! dx \,
{v_\mu \over 2\pi}\left[K_\mu(\partial_x \phi_\mu)^2
\! +\! K^{-1}_\mu(\partial_x \theta_\mu)^2 \right],
\end{equation}
where $K_c<1$ (repulsive interaction)
and $K_s=1$ (spin SU(2) symmetry).
$H_{S}$ is given by $H_{N}+H_\Delta$,
where 
\begin{equation}
H_\Delta =-\Delta\int_{-\infty}^{0}{dx \over x_0} \,
\cos \sqrt{2}\phi_c \cos \sqrt{2} \theta_s,
\end{equation}
is the bosonized expression for
$\int dx \sum_\sigma (\Delta 
\psi^\dagger_{R,\sigma}\psi^\dagger_{L,-\sigma}+{\rm h.c.})$,
and $x_0$ is the short length scale cut-off. 
$\Delta=\Delta(T)$ is assumed to be a real positive constant.
Here $H_N$ is identical to $H_{LL}$,
except that $K_{c,s}=1$ and the integration over $x$ runs from
$-\infty$ to 0. 
Finally the impurity backscattering potential at the interface is 
\begin{equation}
H_{\rm imp}= -U_0 \cos \sqrt{2}\theta_c(x=0) \cos \sqrt{2} \theta_s(x=0).
\end{equation}
A similar 1d model is suggested previously
in Refs.\cite{Maslov96PRB,Affleck00PRB}.

{\it Ideal clean interface}: 
We first consider an ideal interface with $U_0=0$.
Above the critical temperature $T_c$, where $\Delta(T)=0$,
$H_\Delta$ vanishes
and the system reduces to the N-LL junction with an ideal interface.
By using the technique in Ref.~\cite{Kane92PRB},
one can integrate out degrees of freedom for all $x\neq 0$ 
and obtain an imaginary time effective action $S$ for 
$\theta^0_\mu(\tau)\equiv \theta_\mu(x=0,\tau)$,
\begin{equation}
S= \sum_{\omega_n}{|\omega_n| \over \pi \beta}\left[
 {K_c^{-1}+1 \over 2}\ |\theta^0_c(\omega_n)|^2
+|\theta^0_s(\omega_n)|^2\right]\, ,
\label{eq:S-clean-high}
\end{equation}
where $\beta=T^{-1}$ is the inverse temperature, and 
$\theta^0_\mu(\omega)\equiv
\int_0^\beta d\tau e^{i\omega \tau}\theta^0_\mu(\tau)$.
From the correlation function $\left< \theta_c^0(\tau)\theta_c^0(0)\right>$,
one obtains the zero bias conductance 
$G=G_{\rm NL}\equiv 2K_c/(1+K_c)G_0$~\cite{Luttinger-factor} 
where $G_0=2e^2/h$.

On the other hand, for $T\ll T_c$, where 
$\Delta(T)\gg T$,
$H_\Delta$ becomes important and
the cosine functions in $H_\Delta$ can be expanded
around a potential minimum, which
results in mass terms for the fields $\phi_c$ and $\theta_s$.
Since fluctuations of $\phi_c$ and $\theta_c$
are strongly suppressed by the mass terms for $x\le 0$,
$\phi_c^0$ and $\theta_s^0$ may be regarded as constants,
for example, zero. 
In the fermionic description,
$\phi_c^0=\theta_s^0=0$ supplies
the boundary conditions (BC), 
$\psi_{R\uparrow}(0)=-i\psi^\dagger_{L\downarrow}(0)$,
$\psi_{R\downarrow}(0)=+i\psi^\dagger_{L\uparrow}(0)$,
which, combined with $H_{LL}$, define a boundary conformal
field theory in the semi-infinite space $x\ge 0$~\cite{Affleck00PRB}.
As demonstrated in Ref.~\cite{Maslov96PRB},
the BCs imply that
all particles incident on the junction interface
from the LL side are Andreev reflected, 
resulting in a transport of spin singlet Cooper pairs
through the junction interface.

The perfect Andreev reflection is expected to
enhance the conductance.
To evaluate $G$,
we integrate out degrees of freedom for $x\neq 0$
and obtain an imaginary time effective action,
\begin{equation}
S= \sum_{\omega_n}{|\omega_n| \over \pi\beta} \left[
{K_c^{-1} \over 2} |\theta^0_c(\omega_n)|^2
+Q(\omega_n)|\theta^0_s(\omega_n)|^2\right]\; ,
\label{eq:S-clean-low}
\end{equation}
where 
$Q(\omega)\propto \Delta/|\omega|$ for $|\omega|\ll \Delta$.
Fluctuations of $\theta_s^0$ are massive as expected.
On the other hand, fluctuations of $\theta_c^0$ not only remain massless
but are enhanced since the coefficient of $|\theta_c^0|^2$ term is
reduced compared to that in Eq.~(\ref{eq:S-clean-high}).
The enhancement is due to the Andreev process
and leads to the increased conductance
$G=G_{\rm SL}\equiv 2K_c G_0$~\cite{Luttinger-factor},
which is larger than $G_{\rm NL}$ 
by the factor $1+K_c$~\cite{Luttinger-factor},
a generalization of the factor 2 enhancement in a S-N junction
with an ideal interface~\cite{Blonder82PRB}.
We then interpolate $G_{\rm NL}$ and $G_{\rm SL}$
to obtain the temperature dependence of $R=G^{-1}$
(curve A in Fig.~\ref{RG_flow}).

{\it Almost ideal interface}: 
When $U_0$ is not zero but sufficiently small,
the effective actions~(\ref{eq:S-clean-high}) and (\ref{eq:S-clean-low})
are weakly perturbed by $\int_0^\beta d\tau H_{\rm imp}$,
which suppresses fluctuations of $\theta_c^0$ and $\theta_s^0$,
and reduces $G$.
Above $T_c$, this
generates a correction
$\delta G\propto -T^{\alpha_1}$ to $G_{\rm NL}$,
where $\alpha_1=(K_c-1)/(K_c+1)<0$, and
far below $T_c$, 
a correction $\delta G\propto -T^{\alpha_2}$ to $G_{\rm SL}$,
where $\alpha_2=2(K_c-1)<0$. 
The power-laws can be verified by using the method
in Ref.~\cite{Kane92PRB} 
and the latter exponent is previously obtained 
in Refs.~\cite{Takane96JPSJ,Affleck00PRB}.
The $T$-dependence of $R$ can be constructed
from the interpolation (the curve B in Fig.~\ref{RG_flow}).
Note that while $dR/dT<0$ for $T>T_c$ and $T\ll T_c$ due to $\delta G$,
the interpolation requires
$dR/dT$ to be {\it positive} in the intermediate range $T\lesssim T_c$
since $G_{\rm SL},G_{\rm NL} \gg \delta G$
for a sufficiently clean interface
and $G_{\rm SL}>G_{\rm NL}$.
This result can be confirmed
in an alternative way.
Direct calculation of $G$ near $T_c$, 
where both $H_{\rm imp}$ and $H_\Delta$ can be treated perturbatively,
results in $G=G_{\rm NL}+c_{\rm imp} U_0^2 T^{\alpha_{\rm imp}}
+c_\Delta \Delta^2 T^{\alpha_\Delta}$,
where $\alpha_{\rm imp}=\alpha_1$ and the constant $c_{\rm imp}<0$
while $\alpha_\Delta=-2$~\cite{why_two} and $c_\Delta>0$.
Thus for a sufficiently clean interface ($U_0\rightarrow 0$),
the temperature dependence of $G$ is governed by the last term
and $dR/dT>0$ since $c_\Delta>0$~\cite{singular_Delta}.
This $T$-dependence of $R$ for $T\lesssim T_c$ 
has not been revealed in earlier works
and constitutes one of main results of this paper.

For later use, here we define
a crossover temperature $T_{\rm cr}$ ($<T_c$),
above which the superconductivity is dominant
and below which the interaction effect is dominant,
as the temperature where  $dR(T)/dT=0$ (Fig.~\ref{RG_flow}).
We also remark that 
at very low temperatures $T \ll T_{\rm cr}$,
the perturbative results
near this ``ideal-junction'' fixed point (FP) 
are not valid 
since $\delta G$ diverges at zero temperature.
This regime is discussed below.

{\it Poorly transmitting interface}: 
When the bare $U_0=\infty$,
the junction is disconnected and $G=0$.
Although the large bare $U_0$ limit is not our concern,
the diverging correction $\delta G$ for a clean interface
suggests that
this ``disconnected-junction'' FP is a natural candidate 
for the new FP that determines the resistance
of a clean junction for $T\ll T_{\rm cr}$.
At this FP, we may impose the fermionic BC,
$\psi_{L,\sigma}(x=0+)=e^{-i\delta}\psi_{R,\sigma}(x=0+)$,
where $\delta$ is a real constant, 
since all electrons incident on the junction interface
should be normal-reflected. 
Combined with $H_{LL}$, the BC defines
another boundary conformal field theory in $x\ge 0$~\cite{Affleck00PRB}.

When the effective $U_0$ is large but finite,
current flow between the S and the LL can be generated
by tunneling processes,
which can be regarded as perturbations to 
the disconnected-junction FP.
For $T\ll T_{\rm cr}\;(<T_c)$, the single-particle tunneling is
exponentially suppressed.
We thus consider only the Cooper pair tunneling 
$H_t=\sum_\sigma \left[
t\psi^\dagger_{R,\sigma}(0-)\psi^\dagger_{L,-\sigma}(0-)
\psi_{L,-\sigma}(0+)\psi_{R,\sigma}(0+)+{\rm h.c.}\right]$,
which, upon bosonization, becomes
\begin{equation}
H_t=t\cos\sqrt{2}(\phi_{c,LL}^0-\phi_{c,S}^0)
\cos\sqrt{2}(\theta_{s,LL}^0-\theta_{s,S}^0),
\label{eq:pair-tunneling}
\end{equation}
where $\phi_{\mu,LL/S}^0\equiv \phi_\mu(x=0\pm)$
and $\theta_{\mu,LL/S}^0\equiv \theta_\mu(x=0\pm)$.
One then obtains
$G\propto T^{\alpha_3}$, where
$\alpha_3=2(K_c^{-1}-1)>0$.
Note that the power-law dependence justifies
the neglect of the exponentially suppressed single-particle contribution.
The same exponent is previously obtained 
in Refs.~\cite{Takane96JPSJ,Affleck00PRB}.
All curves in Fig.~\ref{RG_flow} for nonzero bare $U_0$
follow the power-law $R\propto T^{-\alpha_3}$  
at sufficiently low temperatures,
where the interaction wins over the superconductivity.

For completeness, we also consider
the temperature dependence above $T_c$
in a junction with sufficiently large {\it bare} $U_0$. 
Above $T_c$, $\Delta=0$ and
the single-particle tunneling leads to
$G\propto T^{\alpha_4}$ where
$\alpha_4=(K_c^{-1}-1)/2>0$.
Note that $\alpha_3$ and $\alpha_4$ are both positive,
implying that the disconnected-junction FP is a stable one.
The curve E in Fig.~\ref{RG_flow}
schematically shows the resistance of
the system when the bare $U_0$ is sufficiently large.
Note that $R$ rises more rapidly below $T_c$
since $\alpha_3>\alpha_4$.

{\it General interface}: To obtain qualitative behavior
for arbitrary bare $U_0$,
we interpolate the two opposite limits 
(curves B and E in Fig.~\ref{RG_flow}).
The interpolation assumes the absence of
a ``partially-transmitting-junction'' FP.
The assumption is supported by a recent work~\cite{Affleck00PRB},
and also justifies the interpolation
to the intermediate temperature range
used to construct the curves in Fig.~\ref{RG_flow}.
A natural interpolation to intermediate $U_0$ is depicted in
Fig.~\ref{RG_flow}:
For nonzero $U_0$ larger than a critical value $U_{0,c}$,
$R$ increases monotonically with lowering temperature (curves D and E)
while for $U_0<U_{0,c}$, 
$R$ shows nonmonotonic behavior,
that is, $dR/dT>0$ for $T_{\rm cr}<T<T_c$
and $dR/dT<0$ for $T<T_{\rm cr}$ (curves B and C).

{\it $dV/dI$--$V$ characteristic}:
Here we study the $dV/dI$--$V$ characteristic
at $T<T_c$ for almost ideal interfaces.
We also confine ourselves to the subgap range $eV<\Delta$.
When $T$ is slightly below $T_c$, 
effects of the superconductor are weak
and the characteristic should be similar
to the characteristic for a N-LL junction,
for which $dV/dI$ decays monotonically with increasing $V$
(curve 10 in Fig.~\ref{R-V}).
As $T$ is lowered towards $T_{\rm cr}$,
effects of the superconductor become stronger.
Near the zero bias,
according to the curves B and C in Fig.~\ref{RG_flow},
the differential resistance
should decay due to the effects of the superconductor. 
At relatively high bias $eV\approx \Delta$,
on the other hand, effects of the superconductor
should remain still weak since the small energy barrier $\Delta-eV$
for single particle transport can be overcome by thermal fluctuations.
As a result of different effectiveness of $\Delta$
at low and high biases,
$dV/dI$ develops a dip at zero bias (curves 9 to 5 in Fig.~\ref{R-V}).
This dip becomes deeper as $T$ is lowered.

When $T$ is lowered below $T_{\rm cr}$,
the large effective $U_0$ due to
the repulsive interaction wins
over the effects of the superconductivity
as shown in the curves B and C in Fig.~\ref{RG_flow}.
Thus near the zero bias, $dV/dI$  should grow
as $T$ approaches zero.
However for sufficiently large bias,
only minor changes in $dV/dI$ are expected
since the bias voltage itself serves as an energy cutoff for
the RG flow of $U_0$ and prohibits the crossover.
As a result, in the temperature range $0<T<T_{\rm cr}$,
the characteristic becomes nonmonotonic in $V$:
$dV/dI$ has a negative slope 
below a certain temperature-dependent crossover bias $V_{\rm cr}(T)$
and a positive slope above $V_{\rm cr}(T)$ (curves 5 to 0 in Fig.~\ref{R-V}).
Here
it is reasonable to expect that
 $V_{\rm cr}(T=T_{\rm cr})=0$ and
$V_{\rm cr}$ grows monotonically to its zero temperature value
as $T\rightarrow 0$.
At $T=0$ (curve 0), 
$R$ follows the power-law $V^{-\alpha_3}$
for $V\ll V_{\rm cr}(0)$.
Figure~\ref{R-V} summarizes the second main result of the paper.

{\it Discussion}: 
The present analysis may be tested with single-wall CNTs.
%
Although a single-wall CNT has {\it two} transport channels rather than one,
it can be analyzed in a similar way
by exploiting the mapping from a single-wall armchair
CNT to a two-leg ladder~\cite{Balents96cond-mat}.
The conductance calculations closely parallel 
those for the single channel system.
For $U_0=0$, $G$ crosses over from
$G_{\rm NL}=4K_{c+}/(1+K_{c+})G_0$ ($T>T_c$) to 
$G_{\rm SL}=4K_{c+}G_0$ ($T<T_c$)~\cite{Luttinger-factor}.
For small $U_0$, the power-law correction to $G$ appears,
whose temperature scaling exponents $\alpha_{1,2}$ have new values,
$\alpha_1=(K_{c+}-1)/2(K_{c+}+1)<0$
and $\alpha_2=K_{c+}-1<0$. 
For (effectively) large $U_0$, the leading contribution to $G$
again scales with temperature but the exponents $\alpha_{3,4}$ are
modified to
$\alpha_3=K_{c+}^{-1}-1>0$ and $\alpha_4=(K_{c+}^{-1}-1)/4>0$.
Then by interpolating the behaviors
in the high and low energy scales,
we find that the results in Fig.~\ref{RG_flow}
remain equally valid for the two-leg ladder system.
It can also be verified that
the $dV/dI$--$V$ characteristic remains qualitatively
similar.

Two recent experiments~\cite{Kasumov99Science,Morpurgo99Science}
reported measurements 
on single-wall CNTs in contact with two superconducting leads.
Thus the experimental configuration amounts to
a S-CNT-S junction rather than a S-LL junction,
and the Josephson current is observed in Ref.~\cite{Kasumov99Science}.
In Ref.~\cite{Morpurgo99Science}, however,
the Josephson current is not observed and
it is suggested that the system for some unknown reason
behaves as an ``incoherent'' series of two S-CNT junctions,
where single S-CNT junction dynamics essentially determines
properties of the whole system.
Motivated by this observation,
we compare our main results (Figs.~\ref{RG_flow}, and \ref{R-V})
with the results in Ref.~\cite{Morpurgo99Science}.
For a sufficiently clean junction interface,
the experiment found that 
the zero bias resistance shows nonmonotonic behavior,
$dR/dT>0$ for $T_{\rm cr}(\approx 4\;{\rm K})<T<T_c(\approx 10\; {\rm K})$
and $dR/dT<0$ for $T<T_{\rm cr}$.
This behavior is in qualitative  agreement 
with the curves B and C in Fig.~\ref{RG_flow}.
The experiment also found that 
the $dV/dI$--$V$ characteristics above and below $T_{\rm cr}$ 
are qualitatively different,
again in qualitative agreement with Fig.~\ref{R-V}.

However it is yet premature to interpret Ref.~\cite{Morpurgo99Science}
as an experimental verification of the predictions for a S-LL junction.
As recently pointed out by Wei {\it et al.}~\cite{Wei01PRB},
the level spacing of CNTs due to the finite size effect
in Ref.~\cite{Morpurgo99Science} 
is of the same order as $\Delta$.
The level spacing can serve as a new energy cut-off
and prevent the system from reaching the
low energy regime in Figs.~\ref{RG_flow} and \ref{R-V}.
Moreover it is demonstrated~\cite{Wei01PRB,Aleiner96PRB}
that a {\it noninteracting} model of a S-CNT-N junction,
which takes into account the finite size effect, 
can produce nonmonotonic dependences.
In this explanation, however,
the resistance does {\it not} follow the power-law 
even sufficiently below $T_{\rm cr}$,
in clear contrast to the predictions for a S-LL junction. 
Thus for an unambiguous verification of the predictions in this paper,
we suggest a search for power-law behaviors
in new experiments with longer CNTs,
where the finite size effect is weaker.
%


{\it Summary}: It is demonstrated that the resistance
of a superconductor-Luttinger liquid junction 
with a sufficiently clean interface
shows nonmonotonic temperature- and voltage-dependence 
due to the competition between 
the superconductivity and the repulsive interaction.
The results are discussed in connection with
a recent experiment~\cite{Morpurgo99Science}
on single-wall carbon nanotubes in contact
with superconductors.

Helpful comments by M.-S. Choi, A. Furusaki, D. Loss, Y. Kiem,
L. Glazman, C. L. Kane, and L. S. Levitov are acknowledged.
HWL and HY were partially supported by 
Swiss-Korean Outstanding Research Efforts Award Program
and HCL and HYC were  supported 
by the Korea Science and Engineering Foundation (KOSEF) 
through the grant No. 1999-2-11400-005-5, and 
by the Ministry of Education through Brain Korea 21 SNU-SKKU Program.
\bibliographystyle{apsrev}

\begin{thebibliography}{99}
\bibitem{Voit94RPP} See for instance J. Voit, Rep. Prog. Phys. 
  {\bf 57}, 977 (1994).
\bibitem{Egger97PRL} R. Egger and A. O. Gogolin, Phys. Rev. Lett.
  {\bf 79}, 5082 (1997); C. Kane {\it et al}., {\it ibid}.  
  {\bf 79}, 5086 (1997).
\bibitem{Bockrath97Science} M. Bockrath {\it et al}., Science
  {\bf 275}, 1922 (1997); M. Bockrath {\it et al}., Nature
  {\bf 397}, 598 (1999); Z. Yao {\it et al}., {\it ibid}.
  {\bf 402}, 273 (1999).
\bibitem{Fazio95PRL} R. Fazio, F. W. J. Hekking, and A. A. Odintsov,
  Phys. Rev. Lett. {\bf 74}, 1843 (1995).
\bibitem{Maslov96PRB} D. L. Maslov {\it et al}.,
  Phys. Rev. B {\bf 53}, 1548 (1996).
\bibitem{Takane96JPSJ} Y. Takane and Y. Koyama, J. Phys. Soc. Jpn.
  {\bf 65}, 3630 (1996); Y. Takane, {\it ibid}. {\bf 66}, 537 (1997).
\bibitem{Affleck00PRB} I. Affleck, J.-S. Caux, and A. M. Zagoskin, 
  Phys. Rev. B {\bf 62}, 1433 (2000).
\bibitem{Caux01cond-mat} J.-S. Caux, H. Saleur, and F. Siano,
  cond-mat/0109103.
\bibitem{Kasumov99Science} A. Yu. Kasumov {\it et al}.,
  Science {\bf 284}, 1508 (1999).
\bibitem{Morpurgo99Science} A. F. Morpurgo {\it et al}.,
  Science {\bf 286}, 263 (1999).
\bibitem{Blonder82PRB} G. E. Blonder, M. Tinkham, and T. M. Klapwijk,
  Phys. Rev. B {\bf 25}, 4515 (1982).
\bibitem{Kane92PRB} C. L. Kane and M. P. A. Fisher,
  Phys. Rev. B {\bf 46}, 15233 (1992);
  A. Furusaki and N. Nagaosa, 
  Phys. Rev. B {\bf 47}, 4631 (1993).
\bibitem{Luttinger-factor} When the LL is connected
  at $x=\infty$ to a normal metal electrode, 
  $K_c$ ($K_{c+}$) that appears in the dc conductance
  of an {\it impurity-free} system 
  is effectively renormalized to one 
  due to the mechanism demonstrated by
  D. L. Maslov and M. Stone [Phys. Rev. B {\bf 52}, 5539 (1995)].
%
\bibitem{why_two} Note that the exponent does not
depend on the Luttinger parameter since $H_S$ is a {\it bulk}
perturbation to $x<0$. The same exponent is obtained
in Ref.~\protect\cite{Blonder82PRB} when $T\gg \Delta$.  
%
\bibitem{singular_Delta} The effect of the last term becomes even stronger
when the singular $T$-dependence of $\Delta \propto (T_c-T)^{1/2}$
near $T_c$ is considered. We note however that $R$ itself may not
show such a singular $T$-dependence  
due to strong fluctuations of $\Delta$ near $T_c$,
which are not taken into account in our model.
%
\bibitem{Balents96cond-mat} L. Balents {\it et al}., cond-mat/9611126.
\bibitem{Wei01PRB} Y. Wei {\it et al}., Phys. Rev. B {\bf 63}, 195412 (2001). 
\bibitem{Aleiner96PRB} I. L. Aleiner, P. Clarke, and L. I. Glazman,
  Phys. Rev. B {\bf 53}, R7630 (1996).
%
\end{thebibliography}


\end{document}